# Signature of spin-resolved quantum point contact in p-type trilayer WSe$_2$ van der Waals heterostructure

*Kohei Sakanashi,[1][*] Peter Krüger,[1] Kenji Watanabe,[2] Takashi Taniguchi,[3]*

*Gil-Ho Kim,[4] David K. Ferry,[5] Jonathan P. Bird,[1,6] and Nobuyuki Aoki[1][*]*

[1]Department of Materials Science, Chiba University, Chiba, 263-8522, Japan

[2]International Center for Materials Nanoartchitectonics, National Institute for Materials Science, Tsukuba, 305-0044, Japan

[3]Research Center for Functional Materials, National Institute for Materials Science, Tsukuba, 305-0044, Japan

[4]School of Electronic and Electrical Engineering and Sungkyunkwan Advanced Institute of Nanotechnology (SAINT), Sungkyunkwan University, Suwon, 16419, South Korea

[5]School of Electrical, Computer and Energy Engineering, Arizona State University, Tempe, Arizona, 85287, U.S.A.

---

[*] Author to whom correspondence should be addressed. E-mail: k.sakanashi@chiba-u.jp, n-aoki@faculty.chiba-u.jp

[6]Department of Electrical Engineering, the State University of New York, University at Buffalo, Buffalo, New York, 14260, U.S.A.

ABSTRACT: In this study, an electrostatically induced quantum confinement structure, so-called quantum point contact, has been realized in a p-type trilayer tungsten diselenide-based van der Waals heterostructure with modified van der Waals contact method with degenerately doped transition metal dichalcogenide crystals. Clear quantized conductance and pinch-off state through the one-dimensional confinement were observed by dual-gating of split gate electrodes and top gate. Conductance plateaus were observed at step of $e^2/h$ in addition to quarter plateaus such as $0.25 \times 2\ e^2/h$ at a finite bias voltage condition indicating the signature of intrinsic spin-polarized quantum point contact.

Quantum devices, such as a quantum point contact (QPC) or a quantum dot (QD) in two-dimensional (2D) material, including graphene and transition metal dichalcogenides (TMDC), have been receiving significant attention as promising materials for future spin-valley coupled qubits[1] or valleytronic device applications such as valley filter.[2,3] Additionally, the strong intrinsic spin-orbit coupling (SOC) in TMDC materials[1] coupled with quantum nanostructure expand the capability of spin-based quantum devices.[4–6] Another possibility of quantum confinement in TMDC offers the confinement of optically excited quasi-particles,[7,8] and the new physics related to exciton condensation can be artificially realized like being trapped in Moiré potentials.[9,10] To date, however, the number of electrostatically induced quantum confinement devices in TMDC or similar 2D materials[7,11–18] was limited due to their poor crystal quality, mobility, ambient stability, and contact properties. Various methods such as doped or gated graphene contact,[7,17] phase engineering,[19,20] transferred-metal top contact,[21,22] or cleaned bottom contact architecture[23–27] assembled in a glove box were reliably realized for TMDC devices although the quality of low-temperature Ohmic contact is still limited. The Ohmic contact method for p-type material like $WSe_2$ was realized via these methods, but it still has high contact resistance and is limited to inert atmosphere assembling.[22,26,27] The graphene contact is a potentially versatile method to achieve Ohmic contact for 2D materials due to their self-assembled clean 2D/2D interface held together by van der Waals force.[17,28] However the method is limited to low-electron-affinity n-type materials such as $MoS_2$ because of the relatively small work function of graphene $\Phi_{WF} \sim 4.1$ eV[29] and the small intrinsic carrier density. Also, the two-dimensional contact by the other branches of metallic 2D materials has been performed while their stability is usually low.[29] Degenerately p-type doped 2D semiconductors with Nb or Re substitution were recently realized,[30] and typically the conductance and carrier density are quite high for the thick crystals. They are thermodynamically-stable degenerately doped semiconductors, and Ohmic contacts have been realized in p-type WSe$_2$ and MoSe$_2$ systems by using these semiconductors as the lead electrodes even at low temperature.[31,32] In the present study, a QPC has been realized in the p-type, high-quality trilayer WSe$_2$, and the spin-resolved transport has been observed within the electrostatically confined nanostructure, which paves the way for realizing the electrostatically controlled quantum nano device coupled with rich physical properties of TMDC such as spin-filter or spin-valley filter devices. Electron transport in a conduction band of bilayer or trilayer TMDC materials are usually thought to happen at the Γ point or at a particular point (Q) on the Γ-K line. However, recent studies predict that the transport is a single layer phenomenon due to the lower tunneling coupling between single layers in 2H stacking by absence of K-K' momentum transfer.[24,33] A similar opportunity of single layer transport would be expected also in the hole transport. In addition, Γ-K transition was observed for trilayer WSe$_2$ due to the perpendicular electric field,[34] therefore the trilayer WSe$_2$ can be a platform for integration of quantum confinement structure and spin-orbit coupling with higher mobility and transparent contacts rather than the monolayer one.

To achieve better contact properties and mobility for trilayer WSe$_2$ crystals, we assembled a van der Waals (vdW) heterostructure of Nb-doped p$^+$-MoS$_2$/WSe$_2$, which is encapsulated between hexagonal boron nitride (hBN) flakes by a standard polymer-based dry transfer technique.[35] The stack was placed on top of pre-fabricated thin Cr/Pd bottom split gates shown in the inset of Figure 1a on a SiO$_2$/Si substrate. The bottom split gates have dimensions of 80 nm wide and 50 nm long. All experiments were done with the right set of the split gates. To form Ohmic contacts for the p$^+$-MoS$_2$ crystals, which act as electrical contacts for the WSe$_2$, Cr/Pd/Au contacts were patterned by means of electron beam lithography, reactive ion etching with CHF$_3$/O$_2$ plasma, and electron beam deposition in the same manner as a one-dimensional edge contact for graphene.[35] Then, the top gate electrode was deposited and used for tuning both contact properties and channel carrier density (see supporting information S1 for details). During all electron-beam lithography processing, we kept the acceleration voltage below 20 kV to minimize the electron beam irradiation damage for TMDC crystals.[36] The optical microscope image and the schematic cross-sectional image of the complete device structure are shown in Figure 1a,b, respectively. All transport measurements were done in a quasi-four-terminal measurement configuration (see supporting information S1 for details) except for the current-voltage (*I-V*) measurement, and performed in JANIS $^3$He insert, which has a base temperature of 300 mK. However, in order to avoid conductance fluctuations due to the disorder related to the intrinsic vacancy, strain induced fluctuation, or charged impurity, almost all of the low temperature measurements were performed at 1.8 K.

Employing 2D/2D contact methods, we achieved an Ohmic behavior at room temperature and low temperature as shown in Figure 2a,b with linear *I-V* characteristics in two-terminal configuration in appropriate gate bias conditions, and observed a metal-insulator transition behavior of conductance properties as a function of the global top gate voltage ($V_{TG}$) as shown in Figure 2c. The fixed point of metal-insulator transition is not well-defined because both the series resistance of the p$^+$-MoS$_2$ crystals and the contact resistance between the p$^+$-MoS$_2$ and the WSe$_2$ have a non-monotonic dependence on temperature. Our device has a four-terminal geometry, however this geometry is not able to remove such series resistances completely (see supporting information S3 for details). Using the quasi four-terminal configuration, the field-effect mobility of the two-dimensional hole gas (2DHG) in the WSe$_2$ region was estimated at ~1000 cm$^2$/Vs at low density (see S3 for details). However, the Shubnikov-de Haas (SdH) oscillations start from roughly 4 Tesla suggesting that the estimated quantum mobility $\mu_Q \sim 1/B_C$ is around 2500 cm$^2$/Vs which is comparable with former studies of pre-fabricated bottom Pt contacts or transferred Pt contacts scheme.[22,26,27] Attributing the quasi 4-terminal resistance values ($R_{4t}$), the contact resistance between $p^+$-$MoS_2$ and $WSe_2$ is calculated from the channel resistivity ($\rho_{ch}$, estimated from the carrier density and the quantum mobility), the series $p^+$-$MoS_2$ resistance (see S3 for details), and the device geometry as $R_C = (R_{4t} - \rho_{ch} \cdot L/W - 2\ R_{p\text{-}MoS2})/2$. The estimated contact resistance at $T = 1.8$ K and $n = -4.1 \times 10^{12}$ cm$^{-2}$ is 2140 Ω which is also comparable with the best value obtained in previous studies without using inert atmosphere and complex device fabrication method.[22,26,27] The estimated mean free path at proper carrier density should be larger than 100 nm for the $WSe_2$ layer, and ballistic or quasi-ballistic transport would be realized at least within the QPC whose channel length was 50 nm long. The beating of SdH oscillations shown in Figure 2d may be attributed to the subband splitting, and doubling of Fermi surface due to their spin-polarized state, or to the presence of two contributions to the carrier density at Γ and K point in valence band (see S4 for details). The 2DHG carrier density was set to slightly more than $3.0 \times 10^{12}$ cm$^{-2}$ ($V_{TG} < -11$ V) in order to achieve Ohmic contacts and to reduce the localization effect due to disorder (see also S3). Two stages of transition in the conductance curves are observed in Figure 3a. By applying a positive split gate voltage ($V_{SG}$), the conductance begins to decrease since the 2DHG above the split gates has started to deplete (the first stage transition). When the 2DHG is depleted completely, a narrow constriction was formed and the conductance curve begins to decrease until the pinched-off state by squeezing the channel width of the constriction. After subtraction of the series resistance $R_S = R_{4t}$ ($V_{SG} = 0$) from each curve, multiple quantized conductance steps appear below 4 $e^2/h$ at 1.8 K as shown in Figure 3b. The clear plateau features are observed to be quantized in a unit of $e^2/h$ instead of 2 $e^2/h$ with spin degeneracy for all curves. Similar quantized conductance steps were also observed in another set of split gates although some resonant peaks were included due to presence of localized states in the channel region. Higher carrier density usually provides higher carrier mobility in TMDC devices, however application of a much higher $V_{SG}$ value is needed to achieve the pinch-off of current. Therefore, we set the $V_{TG}$ range as $-13$ V $\leq V_{TG} \leq -11$ V during the QPC operations to maintain the gate leakage current in the pico-amp range. Figure 3c shows plots of the conductance values with respect to the derivative of the conductance, $dG/dV_{SG}$. We can confirm that application of lower $V_{TG}$ such as $-13$ V is not appropriate for observing more than three clear quantized conductance steps. To examine the nature of $e^2/h$ steps at zero magnetic field, a perpendicular magnetic field was applied and the transmission curves were observed at every 0.6 T as shown in Figure 3d. All the curves are corrected by subtracting experimentally obtained $R_{4t}$ value at $V_{SG} = 0$ V in each curve as the series resistance in the same manner at zero magnetic field. The conductance curves under the magnetic field showed well-resolved quantized conductance steps although the conductance values at the steps are slightly deviated from $e^2/h$ as the magnetic field becomes larger especially more than 4 T and at the higher conductance stepes. Some of the reasons for the deviation can be considered, and they may appear at the same time. First, we need to care about spin density theory, in which the plateaus value depends on the spin-density in self-energy correction in the QPC conduction. Also the magnetic field provides a depopulation effect that changes the spin density in the QPC channel. The QPC plateaus should be shifted from any integer number plateau such as $N \times e^2/h$ (where $N$ is an integer number).[37] The other possibility is the effect of quantum Hall effect under perpendicular magnetic field. The tunneling coupling between the quantum Hall edge channels in the 2DHG and the one-dimensional channel in such a narrow QPC may provide additional series resistance that has not been cared at the case without forming the QPC at $V_{SG} = 0$ V. Moreover, due to the quasi-four terminal measurement configurations in our device, the series resistance must include both components of the oscillation from Shubnikov-de Haas oscillation and the Hall resistance. By subtracting such additional series resistance, the plateau values may be corrected appropriately however there is no way to estimate the quantitative value of such effects and eliminate it correctly from the conductance curves of the QPC.[38,39] Nevertheless, the correction of series resistance is necessary because the series resistance is the same order of the QPC resistance itself in this device. Therefore, we decided to correct the conductance curves by subtracting the experimental resistance value of $R_{4t}$ at $V_{SG} = 0$ V at every magnetic field. However, it is clear that the differential conductance lines $dG/dV_{SG}$ grow successively until 9 T without splitting or crossing each other as shown in Figure 3e. These results may indicate that the spin has already been polarized even at zero magnetic field although additional experiments and theoretical calculations are necessary to explore the mechanisms of this phenomenon.

The degeneracy of holes in trilayer $WSe_2$ is two when assuming a spin degeneracy $g_s = 2$ and valley degeneracy $g_v = 1$. This is suggested by DFT calculations of trilayer $WSe_2$ in absence of an electric field. In this case, the valence band maximum is at the Γ point, which is spin-degenerate by symmetry (see S5 for details). However, the valence band edge at the K point has a spin-polarized band splitting as large as ~30 meV induced by the spin-orbit coupling. In addition, application of a strong perpendicular electric field $D = (\varepsilon_{TG}V_{TG}/d_{TG} - \varepsilon_{SG}V_{SG}/d_{SG})/2$ where $\varepsilon_{TG}$ ($\varepsilon_{SG}$) is the dielectric constant of the top (bottom) hBN layers of thickness $d_{TG}$ ($d_{SG}$), lifts the band edge at the K point.[34,40,41] Our DFT calculations confirm that the valence band edge switches from Γ to K for $D > 1$ V/nm (see S5 for details). Looking at the K point maxima of the valence band under the $D$ field, the spin-polarized bands along a perpendicular direction are still present at the K point even under $D$ field condition and the energy split becomes more resolved. Under the electric field condition, another degree of freedom, K and K' point valley, is assumed to behave as monolayer materials while the symmetry breaking of valley degeneracy is predicted for K or Q valley of trilayer materials due to the opposite polarity of non-trivial Berry curvature for K and K' valleys.[42] The *D* value induced above the split gates region is estimated at more than 0.75 V/ nm. However, the stray *D* field at the channel region between the split gates must be smaller than the value on the split gates and thus it is insufficient to lift the K point fully to the valence band maxima in our device configurations. Nevertheless, our results suggest a significant contribution of the K point to the spin-resolved transport. Recently, a monolayer like transport behavior has been observed in bilayer or trilayer $MoS_2$ with a dual gate architecture and caused by single layer charge accumulation and screening effect measured by compressibility.[24,33] Such a behavior has been observed in high quality few-layer devices, and might be due to the formation of a quantum well state because of the very weak tunnel coupling between the adjacent layers because of an absence of K-K' transitions in the 2H stacked TMDC material. It is reasonable to consider that the signature of spin-resolved transport observed in our p-type trilayer $WSe_2$-QPC device occurred through the single layer closest to the top gate contributing to the valence band edge at K point.

To examine the intrinsic spin splitting state even at zero magnetic field, we have performed finite source drain bias spectroscopy measurements as a function of $V_{SG}$ at 1.8 K for zero and finite magnetic field conditions as shown in Figure 4. In the finite direct-current (DC) bias spectroscopy, the applied DC voltage is dissipated also by the series resistance, therefore the true applied voltage across QPC ($V_{QPC}$) was appropriately corrected based on principles of voltage distribution (see S6 for details).[43] For the conductance value, the effect has been corrected using those values at the zero bias condition for each curve. Therefore, at least the conductance plateaus around zero-bias voltage can be confirmed to be situated at 0.5 $G_0$, 1.0 $G_0$ and 1.5 $G_0$, where $G_0 = 2\ e^2/h$ and dense-lines regions correspond to conductance plateaus in the transmission curves. Extensive half plateaus in QPC corresponding to a spin splitting state should appear at the conductance of ($G_0 \times N \pm 0.5$)/2, or as 0.25 $G_0$ or 0.75 $G_0$ in the finite bias $V_{QPC}$ instead of $G_0 \times N \pm 0.5$, or as 0.5 $G_0$ or 1.5 $G_0$ as normal spin degenerated states usually observed in HEMT systems. In Figure 4a, clear half plateau states can be observed as dense lines at 0.25 $G_0$ in both bias directions. Although some higher half-plateau states can be confirmed at finite bias shown as the region of dense lines marked by the red-dotted circles, we cannot assign the exact conductance values except 0.25 $G_0$. This is because the conductance values appear lower than the true ones except around the zero-bias region because the correction of non-linear series resistance fails at biased condition coming from low device quality comparing to conventional HEMT system, higher scattering process in QPC channel, the drift of local potential and measurement system by long measurement time of 15 hours,[44] or conductance decreasing or increasing by self-gating effects in the QPC channel.[45,46] However, it is clear that these spin-polarized quarter plateaus correspond to the dark spots which appear at finite bias region in the contour plots of the spectroscopy maps show in Figure S10 of the Supporting Information. The energy spacing between subbands can be estimated by the distance between the half-plateaus. For the $N = 1$ mode, a spacing of approximately 2.7 ± 0.25 meV was found which is comparable to the thermal energy at threshold temperature, ~30 K, at which a conductance step appears in the QPC operations[44] (see S7 and S8 for details). The half plateaus suggesting the spin resolved state at 0.25 $G_0$ are still clearly visible in the presence of a finite magnetic field below the onset of SdH oscillations. At higher magnetic field, the zero bias peaks around $N = 2$ modes split into two peaks as shown in Figures 4b-d even though the visibility of the peak splitting is quite weak at zero magnetic field. However, the splitting of the zero bias peak may be attributed to several mechanisms. One of the promising models is by the two-impurity Kondo effect of two-localized spins in the QPC.[47–50] In this scenario, a spin-polarization of the QPC subbands may arise even at zero magnetic field because these subbands are formed at the spin-polarized K point of the valence band. Such a situation could well match the model of the two-impurity Kondo effect.

In conclusion, the modified 2D/2D contact method allows us to investigate the quantum transport phenomena in the nanostructure fabricated in the p-type trilayer WSe$_2$ with electrostatic gating although the quantum mobility is still limited to ~2500 cm$^2$/Vs. We have realized an electrostatically induced QPC with ~ 2.7 meV subband spacing. Multiple conductance plateaus have been clearly observed and they are quantized in units of $e^2/h$ rather than the spin degenerate value $2e^2/h$. It suggests the existence of the spin-polarized state due to the intrinsic strong SOC at zero magnetic field and 0.25 $G_0$ at biased condition, while the additional experiments and theoretical supports are necessary to explore the nature of spin-resolved transport signature. These results are tentatively attributed to the single-layer transport at K point in the valence band of the trilayer WSe$_2$ within the QPC channel. Our results extend the range of quantum confinement studies down to monolayer with additional valley degeneracy. Also, the use of a broad type of 2D metal, e.g. superconducting NbSe$_2$[29] contact should make it possible to investigate the relation between proximity induced superconducting physics in nanostructures with artificial quantum confinement and strong SOC.

ASSOCIATED CONTENT

**Supporting Information**.

The following files are available free of charge.

Details on fabrication and measurement method, contact $p^+$-$MoS_2$ properties, global top gate voltage dependence of mobility, 2DHG properties, and contact resistance, correction method of DC bias measurements, temperature dependence of QPC property, and color maps of differential conductance in DC bias spectroscopy. (PDF)

AUTHOR INFORMATION

**Author Contributions**

K.S. and N.A. conceived the study; K.S. fabricated the devices and performed the low temperature transport measurement; P.K. performed the density functional theory calculation for band structures; K.S. analyzed data; K.W and T.T grow the high-quality hBN crystals and provided; G.-H.K., D.K.F. and J.P.B. gives theoretical explanation of QPC; K.S., J.P.B., and N.A. wrote the manuscript; and all of authors attend to the discussions.

**Notes**

The authors declare no competing financial interest.

**ACKNOWLEDGMENT**

The authors appreciate the fruitful discussion on materials with Kosuke Nagashio, and thank Takahiro Karatsu and Yewon Kim for their technical assistance of figure preparation. This work was supported by the JSPS KAKENHI Grant Numbers 18H01812, 20J20052, Bilateral Joint Research Project between JSPS and NRF, and CASIO Science Promotion Foundation. G.-H. Kim was supported under the framework of international cooperation program managed by the National Research Foundation of Korea (No. 2021K2A9A2A08000168).

ORCID iDs

Kohei Sakanashi https://orcid.org/0000-0001-6186-0627

Peter Krüger https://orcid.org/0000-0002-1247-9886

Kenji Watanabe https://orcid.org/0000-0003-3701-8119

Takashi Taniguchi https://orcid.org/0000-0002-1467-3105

Gil-Ho Kim https://orcid.org/0000-0002-5153-4235

David K. Ferry https://orcid.org/0000-0002-0942-8033

Jonathan P. Bird https://orcid.org/0000-0002-6966-9007

Nobuyuki Aoki https://orcid.org/0000-0001-9203-6040

**REFERENCES**

(1)  Kormányos, A.; Zólyomi, V.; Drummond, N. D.; Burkard, G. Spin-Orbit Coupling, Quantum Dots, and Qubits in Monolayer Transition Metal Dichalcogenides. *Phys. Rev. X* **2014**, *4* (1), 011034. https://doi.org/10.1103/PhysRevX.4.011034.

(2)  Xiao, D.; Liu, G. Bin; Feng, W.; Xu, X.; Yao, W. Coupled Spin and Valley Physics in Monolayers of $MoS_2$ and Other Group-VI Dichalcogenides. *Phys. Rev. Lett.* **2012**, *108* (19), 196802. https://doi.org/10.1103/PhysRevLett.108.196802.

(3)  Rycerz, A.; Tworzydło, J.; Beenakker, C. W. J. Valley Filter and Valley Valve in Graphene. *Nat. Phys.* **2007**, *3* (3), 172–175. https://doi.org/10.1038/nphys547.

(4)  Frolov, S. M.; Lüscher, S.; Yu, W.; Ren, Y.; Folk, J. A.; Wegscheider, W. Ballistic Spin Resonance. *Nature* **2009**, *458* (7240), 868–871. https://doi.org/10.1038/nature07873.

(5)  Kohda, M.; Okayasu, T.; Nitta, J. Spin-Momentum Locked Spin Manipulation in a Two-Dimensional Rashba System. *Sci. Rep.* **2019**, *9* (1), 1–9. https://doi.org/10.1038/s41598-018-37967-9.

(6) Debray, P.; Rahman, S. M. S.; Wan, J.; Newrock, R. S.; Cahay, M.; Ngo, A. T.; Ulloa, S. E.; Herbert, S. T.; Muhammad, M.; Johnson, M. All-Electric Quantum Point Contact Spin-Polarizer. *Nat. Nanotechnol.* **2009**, *4* (11), 759–764. https://doi.org/10.1038/nnano.2009.240.

(7) Wang, K.; De Greve, K.; Jauregui, L. A.; Sushko, A.; High, A.; Zhou, Y.; Scuri, G.; Taniguchi, T.; Watanabe, K.; Lukin, M. D.; Park, H.; Kim, P. Electrical Control of Charged Carriers and Excitons in Atomically Thin Materials. *Nat. Nanotechnol.* **2018**, *13* (2), 128–132. https://doi.org/10.1038/s41565-017-0030-x.

(8) Xu, C.; Leonard, J. R.; Dorow, C. J.; Butov, L. V.; Fogler, M. M.; Nikonov, D. E.; Young, I. A. Exciton Gas Transport through Nanoconstrictions. *Nano Lett.* **2019**, *19* (8), 5373–5379. https://doi.org/10.1021/acs.nanolett.9b01877.

(9) Wang, Z.; Rhodes, D. A.; Watanabe, K.; Taniguchi, T.; Hone, J. C.; Shan, J.; Mak, K. F. Evidence of High-Temperature Exciton Condensation in Two-Dimensional Atomic Double Layers. *Nature* **2019**, *574* (7776), 76–80. https://doi.org/10.1038/s41586-019-1591-7.

(10) Tran, K.; Moody, G.; Wu, F.; Lu, X.; Choi, J.; Kim, K.; Rai, A.; Sanchez, D. A.; Quan, J.; Singh, A.; Embley, J.; Zepeda, A.; Campbell, M.; Autry, T.; Taniguchi, T.; Watanabe, K.; Lu, N.; Banerjee, S. K.; Silverman, K. L.; Kim, S.; Tutuc, E.; Yang, L.; MacDonald, A. H.; Li, X. Evidence for Moiré Excitons in van Der Waals Heterostructures. *Nature* **2019**, *567* (7746), 71–75. https://doi.org/10.1038/s41586-019-0975-z.

(11) Marinov, K.; Avsar, A.; Watanabe, K.; Taniguchi, T.; Kis, A. Resolving the Spin Splitting in the Conduction Band of Monolayer $MoS_2$. *Nat. Commun.* **2017**, *8* (1), 1–7. https://doi.org/10.1038/s41467-017-02047-5.

(12) Davari, S.; Stacy, J.; Mercado, A. M.; Tull, J. D.; Basnet, R.; Pandey, K.; Watanabe, K.; Taniguchi, T.; Hu, J.; Churchill, H. O. H. Gate-Defined Accumulation-Mode Quantum Dots in Monolayer and Bilayer $WSe_2$. *Phys. Rev. Appl.* **2020**, *13* (5), 054058. https://doi.org/10.1103/PhysRevApplied.13.054058.

(13) Epping, A.; Banszerus, L.; Güttinger, J.; Krückeberg, L.; Watanabe, K.; Taniguchi, T.; Hassler, F.; Beschoten, B.; Stampfer, C. Quantum Transport through $MoS_2$ Constrictions Defined by Photodoping. *J. Phys. Condens. Matter* **2018**, *30* (20), 205001. https://doi.org/10.1088/1361-648X/aabbb8.

(14) Song, X. X.; Liu, D.; Mosallanejad, V.; You, J.; Han, T. Y.; Chen, D. T.; Li, H. O.; Cao, G.; Xiao, M.; Guo, G. C.; Guo, G. P. A Gate Defined Quantum Dot on the Two-Dimensional Transition Metal Dichalcogenide Semiconductor $WSe_2$. *Nanoscale* **2015**, *7* (40), 16867–16873. https://doi.org/10.1039/c5nr04961j.

(15) Song, X. X.; Zhang, Z. Z.; You, J.; Liu, D.; Li, H. O.; Cao, G.; Xiao, M.; Guo, G. P. Temperature Dependence of Coulomb Oscillations in a Few-Layer Two-Dimensional $WS_2$ Quantum Dot. *Sci. Rep.* **2015**, *5*. https://doi.org/10.1038/srep16113.

(16) Hamer, M.; Tóvári, E.; Zhu, M.; Thompson, M. D.; Mayorov, A.; Prance, J.; Lee, Y.; Haley, R. P.; Kudrynskyi, Z. R.; Patanè, A.; Terry, D.; Kovalyuk, Z. D.; Ensslin, K.; Kretinin, A. V.; Geim, A.; Gorbachev, R. Gate-Defined Quantum Confinement in InSe-Based van Der Waals Heterostructures. *Nano Lett.* **2018**, *18* (6), 3950–3955. https://doi.org/10.1021/acs.nanolett.8b01376.

(17) Pisoni, R.; Lee, Y.; Overweg, H.; Eich, M.; Simonet, P.; Watanabe, K.; Taniguchi, T.; Gorbachev, R.; Ihn, T.; Ensslin, K. Gate-Defined One-Dimensional Channel and Broken Symmetry States in $MoS_2$ van Der Waals Heterostructures. *Nano Lett.* **2017**, *17* (8), 5008–5011. https://doi.org/10.1021/acs.nanolett.7b02186.

(18) Pisoni, R.; Lei, Z.; Back, P.; Eich, M.; Overweg, H.; Lee, Y.; Watanabe, K.; Taniguchi, T.; Ihn, T.; Ensslin, K. Gate-Tunable Quantum Dot in a High Quality Single Layer $MoS_2$ van Der Waals Heterostructure. *Appl. Phys. Lett.* **2018**, *112* (12), 123101. https://doi.org/10.1063/1.5021113.

(19) Zhu, J.; Wang, Z.; Yu, H.; Li, N.; Zhang, J.; Meng, J.; Liao, M.; Zhao, J.; Lu, X.; Du, L.; Yang, R.; Shi, D.; Jiang, Y.; Zhang, G. Argon Plasma Induced Phase Transition in Monolayer $MoS_2$. *J. Am. Chem. Soc.* **2017**, *139* (30), 10216–10219. https://doi.org/10.1021/jacs.7b05765.

(20) Sakanashi, K.; Ouchi, H.; Kamiya, K.; Krüger, P.; Miyamoto, K.; Omatsu, T.; Ueno, K.; Watanabe, K.; Taniguchi, T.; Bird, J. P.; Aoki, N. Investigation of Laser-Induced-Metal Phase of MoTe$_2$ and Its Contact Property via Scanning Gate Microscopy. *Nanotechnology* **2020**, *31* (20), 205205. https://doi.org/10.1088/1361-6528/ab71b8.

(21) Telford, E. J.; Benyamini, A.; Rhodes, D.; Wang, D.; Jung, Y.; Zangiabadi, A.; Watanabe, K.; Taniguchi, T.; Jia, S.; Barmak, K.; Pasupathy, A. N.; Dean, C. R.; Hone, J. Via Method for Lithography Free Contact and Preservation of 2D Materials. *Nano Lett.* **2018**, *18* (2), 1416–1420. https://doi.org/10.1021/acs.nanolett.7b05161.

(22) Jung, Y.; Choi, M. S.; Nipane, A.; Borah, A.; Kim, B.; Zangiabadi, A.; Taniguchi, T.; Watanabe, K.; Yoo, W. J.; Hone, J.; Teherani, J. T. Transferred via Contacts as a Platform for Ideal Two-Dimensional Transistors. *Nat. Electron.* **2019**, *2* (5), 187–194. https://doi.org/10.1038/s41928-019-0245-y.

(23) Shi, Q.; Shih, E. M.; Gustafsson, M. V.; Rhodes, D. A.; Kim, B.; Watanabe, K.; Taniguchi, T.; Papić, Z.; Hone, J.; Dean, C. R. Odd- and Even-Denominator Fractional Quantum Hall States in Monolayer WSe$_2$. *Nat. Nanotechnol.* **2020**, *15* (7), 569–573. https://doi.org/10.1038/s41565-020-0685-6.

(24) Pisoni, R.; Davatz, T.; Watanabe, K.; Taniguchi, T.; Ihn, T.; Ensslin, K. Absence of Interlayer Tunnel Coupling of K-Valley Electrons in Bilayer MoS$_2$. *Phys. Rev. Lett.* **2019**, *123* (11), 117702. https://doi.org/10.1103/PhysRevLett.123.117702.

(25) Pisoni, R.; Kormányos, A.; Brooks, M.; Lei, Z.; Back, P.; Eich, M.; Overweg, H.; Lee, Y.; Rickhaus, P.; Watanabe, K.; Taniguchi, T.; Imamoglu, A.; Burkard, G.; Ihn, T.; Ensslin, K. Interactions and Magnetotransport through Spin-Valley Coupled Landau Levels in Monolayer MoS$_2$. *Phys. Rev. Lett.* **2018**, *121* (24), 247701. https://doi.org/10.1103/PhysRevLett.121.247701.

(26) Fallahazad, B.; Movva, H. C. P.; Kim, K.; Larentis, S.; Taniguchi, T.; Watanabe, K.; Banerjee, S. K.; Tutuc, E. Shubnikov-de Haas Oscillations of High-Mobility Holes in Monolayer and Bilayer WSe$_2$: Landau Level Degeneracy, Effective Mass, and Negative

Compressibility. *Phys. Rev. Lett.* **2016**, *116* (8), 086601. https://doi.org/10.1103/PhysRevLett.116.086601.

(27) Movva, H. C. P.; Rai, A.; Kang, S.; Kim, K.; Fallahazad, B.; Taniguchi, T.; Watanabe, K.; Tutuc, E.; Banerjee, S. K. High-Mobility Holes in Dual-Gated WSe$_2$ Field-Effect Transistors. *ACS Nano* **2015**, *9* (10), 10402–10410. https://doi.org/10.1021/acsnano.5b04611.

(28) Cui, X.; Lee, G. H.; Kim, Y. D.; Arefe, G.; Huang, P. Y.; Lee, C. H.; Chenet, D. A.; Zhang, X.; Wang, L.; Ye, F.; Pizzocchero, F.; Jessen, B. S.; Watanabe, K.; Taniguchi, T.; Muller, D. A.; Low, T.; Kim, P.; Hone, J. Multi-Terminal Transport Measurements of MoS$_2$ Using a van Der Waals Heterostructure Device Platform. *Nat. Nanotechnol.* **2015**, *10* (6), 534–540. https://doi.org/10.1038/nnano.2015.70.

(29) Sata, Y.; Moriya, R.; Masubuchi, S.; Watanabe, K.; Taniguchi, T.; Machida, T. N- and p-Type Carrier Injections into WSe$_2$ with van Der Waals Contacts of Two-Dimensional Materials. In *Japanese Journal of Applied Physics*; Japan Society of Applied Physics, 2017; Vol. 56, p 04CK09. https://doi.org/10.7567/JJAP.56.04CK09.

(30) Fang, N.; Toyoda, S.; Taniguchi, T.; Watanabe, K.; Nagashio, K. Full Energy Spectra of Interface State Densities for $n$ - and $p$ -type MoS$_2$ Field-Effect Transistors. *Adv. Funct. Mater.* **2019**, *29* (49), 1904465. https://doi.org/10.1002/adfm.201904465.

(31) Chuang, H. J.; Chamlagain, B.; Koehler, M.; Perera, M. M.; Yan, J.; Mandrus, D.; Tománek, D.; Zhou, Z. Low-Resistance 2D/2D Ohmic Contacts: A Universal Approach to High-Performance WSe$_2$, MoS$_2$, and MoSe$_2$ Transistors. *Nano Lett.* **2016**, *16* (3), 1896–1902. https://doi.org/10.1021/acs.nanolett.5b05066.

(32) Takeyama, K.; Moriya, R.; Watanabe, K.; Masubuchi, S.; Taniguchi, T.; MacHida, T. Low-Temperature p-Type Ohmic Contact to WSe$_2$ using P$^+$-MoS$_2$/WSe$_2$ van Der Waals Interface. *Appl. Phys. Lett.* **2020**, *117* (15), 153101. https://doi.org/10.1063/5.0016468.

(33) Masseroni, M.; Davatz, T.; Pisoni, R.; de Vries, F. K.; Rickhaus, P.; Taniguchi, T.; Watanabe, K.; Falko, V.; Ihn, T.; Ensslin, K. Electron Transport in Dual-Gated Three-Layer MoS$_2$. *Phys. Rev. Res.* **2021**, *3* (2), 23047. https://doi.org/10.1103/PhysRevResearch.3.023047.

(34) Movva, H. C. P.; Lovorn, T.; Fallahazad, B.; Larentis, S.; Kim, K.; Taniguchi, T.; Watanabe, K.; Banerjee, S. K.; MacDonald, A. H.; Tutuc, E. Tunable Γ-K Valley Populations in Hole-Doped Trilayer WSe$_2$. *Phys. Rev. Lett.* **2018**, *120* (10), 107703. https://doi.org/10.1103/PhysRevLett.120.107703.

(35) Wang, L.; Meric, I.; Huang, P. Y.; Gao, Q.; Gao, Y.; Tran, H.; Taniguchi, T.; Watanabe, K.; Campos, L. M.; Muller, D. A.; Guo, J.; Kim, P.; Hone, J.; Shepard, K. L.; Dean, C. R. One-Dimensional Electrical Contact to a Two-Dimensional Material. *Science (80-. ).* **2013**, *342* (6158), 614–617. https://doi.org/10.1126/science.1244358.

(36) Matsunaga, M.; Higuchi, A.; He, G.; Yamada, T.; Krüger, P.; Ochiai, Y.; Gong, Y.; Vajtai, R.; Ajayan, P. M.; Bird, J. P.; Aoki, N. Nanoscale-Barrier Formation Induced by Low-Dose Electron-Beam Exposure in Ultrathin MoS$_2$ Transistors. *ACS Nano* **2016**, *10* (10), 9730–9737. https://doi.org/10.1021/acsnano.6b05952.

(37) Akis, R.; Ferry, D. K. Simulations of Spin Filtering Effects in a Quantum Point Contact. *J. Phys. Condens. Matter* **2008**, *20* (16), 164201. https://doi.org/10.1088/0953-8984/20/16/164201.

(38) Van Wees, B. J.; Kouwenhoven, L. P.; Willems, E. M. M.; Harmans, C. J. P. M.; Mooij, J. E.; Van Houten, H.; Beenakker, C. W. J.; Williamson, J. G.; Foxon, C. T. Quantum Ballistic and Adiabatic Electron Transport Studied with Quantum Point Contacts. *Phys. Rev. B* **1991**, *43* (15), 12431–12453. https://doi.org/10.1103/PhysRevB.43.12431.

(39) Van Wees, B. J.; Kouwenhoven, L. P.; Van Houten, H.; Beenakker, C. W. J.; Mooij, J. E.; Foxon, C. T.; Harris, J. J. Quantized Conductance of Magnetoelectric Subbands in Ballistic Point Contacts. *Phys. Rev. B* **1988**, *38* (5), 3625–3627. https://doi.org/10.1103/PhysRevB.38.3625.

(40) Dai, X.; Li, W.; Wang, T.; Wang, X.; Zhai, C. Bandstructure Modulation of Two-Dimensional WSe$_2$ by Electric Field. *J. Appl. Phys.* **2015**, *117* (8), 084310. https://doi.org/10.1063/1.4907315.

(41) Yuan, H.; Bahramy, M. S.; Morimoto, K.; Wu, S.; Nomura, K.; Yang, B. J.; Shimotani, H.; Suzuki, R.; Toh, M.; Kloc, C.; Xu, X.; Arita, R.; Nagaosa, N.; Iwasa, Y. Zeeman-Type Spin Splitting Controlled by an Electric Field. *Nat. Phys.* **2013**, *9* (9), 563–569. https://doi.org/10.1038/nphys2691.

(42) Wu, Z.; Zhou, B. T.; Cai, X.; Cheung, P.; Liu, G. Bin; Huang, M.; Lin, J.; Han, T.; An, L.; Wang, Y.; Xu, S.; Long, G.; Cheng, C.; Law, K. T.; Zhang, F.; Wang, N. Intrinsic Valley Hall Transport in Atomically Thin MoS$_2$. *Nat. Commun.* **2019**, *10* (1), 1–8. https://doi.org/10.1038/s41467-019-08629-9.

(43) Geier, M.; Freudenfeld, J.; Silva, J. T.; Umansky, V.; Reuter, D.; Wieck, A. D.; Brouwer, P. W.; Ludwig, S. Electrostatic Potential Shape of Gate-Defined Quantum Point Contacts. *Phys. Rev. B* **2020**, *101* (16), 165429. https://doi.org/10.1103/PhysRevB.101.165429.

(44) Rössler, C.; Baer, S.; De Wiljes, E.; Ardelt, P. L.; Ihn, T.; Ensslin, K.; Reichl, C.; Wegscheider, W. Transport Properties of Clean Quantum Point Contacts. *New J. Phys.* **2011**, *13* (11), 113006. https://doi.org/10.1088/1367-2630/13/11/113006.

(45) Kristensen, A.; Bruus, H.; Hansen, A. E.; Jensen, J. B.; Lindelof, P. E.; Marckmann, C. J.; Nyg\aard, J.; S\orensen, C. B.; Beuscher, F.; Forchel, A.; Michel, M. Bias and Temperature Dependence of the 0.7 Conductance Anomaly in Quantum Point Contacts. *Phys. Rev. B* **2000**, *62* (16), 10950–10957. https://doi.org/10.1103/PhysRevB.62.10950.

(46) Patel, N. K.; Nicholls, J. T.; Martn-Moreno, L.; Pepper, M.; Frost, J. E. F.; Ritchie, D. A.; Jones, G. A. C. Evolution of Half Plateaus as a Function of Electric Field in a Ballistic Quasi-One-Dimensional Constriction. *Phys. Rev. B* **1991**, *44* (24), 13549–13555. https://doi.org/10.1103/PhysRevB.44.13549.

(47) Sarkozy, S.; Sfigakis, F.; Das Gupta, K.; Farrer, I.; Ritchie, D. A.; Jones, G. A. C.; Pepper, M. Zero-Bias Anomaly in Quantum Wires. *Phys. Rev. B - Condens. Matter Mater. Phys.* **2009**, *79* (16), 161307. https://doi.org/10.1103/PhysRevB.79.161307.

(48) Liu, T. M.; Hemingway, B.; Kogan, A.; Herbert, S.; Melloch, M. Magnetic Splitting of the Zero-Bias Peak in a Quantum Point Contact with a Tunable Aspect Ratio. *Phys. Rev. B - Condens. Matter Mater. Phys.* **2011**, *84* (7), 075320. https://doi.org/10.1103/PhysRevB.84.075320.

(49) Komijani, Y.; Csontos, M.; Ihn, T.; Ensslin, K.; Meir, Y.; Reuter, D.; Wieck, A. D. Origins of Conductance Anomalies in a P-Type GaAs Quantum Point Contact. *Phys. Rev. B - Condens. Matter Mater. Phys.* **2013**, *87* (24), 245406. https://doi.org/10.1103/PhysRevB.87.245406.

(50) Rokhinson, L. P.; Pfeiffer, L. N.; West, K. W. Spontaneous Spin Polarization in Quantum Point Contacts. *Phys. Rev. Lett.* **2006**, *96* (15), 156602. https://doi.org/10.1103/PhysRevLett.96.156602.

SYNOPSIS TOC

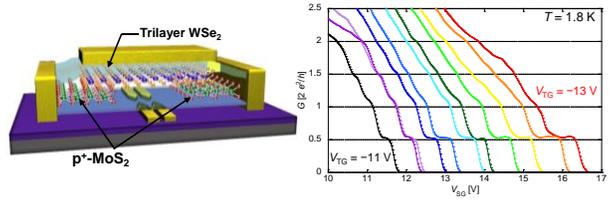

(Half column)

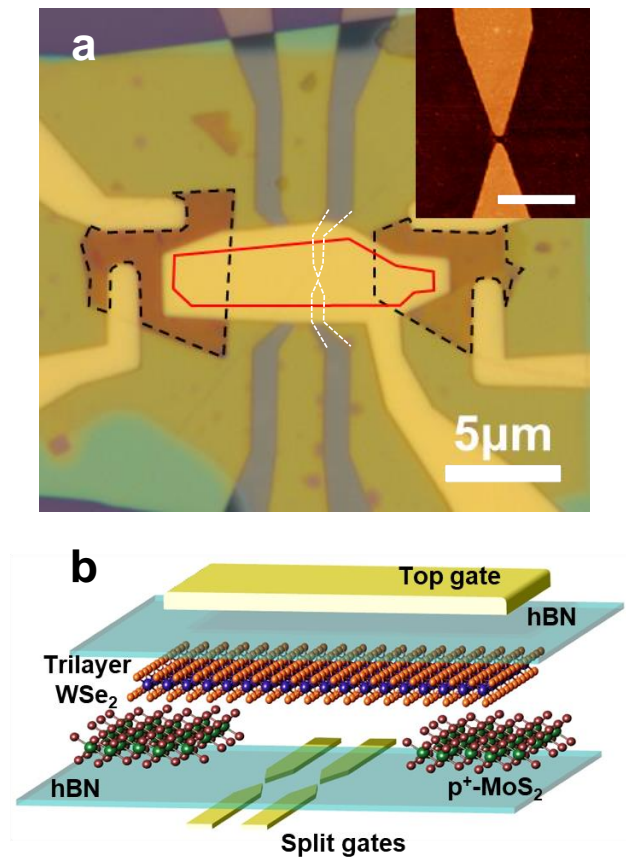

Figure 1. (a) The optical microscope image of WSe$_2$-QPC device. Red solid and black dashed lines indicate the position of trilayer WSe$_2$ and p$^+$-MoS$_2$ crystals, respectively. All flakes are encapsulated by two hBN flakes. Inset: AFM topography image of prefabricated split gates on top of SiO$_2$/Si chip which is illustrated schematically in Figure 1a by white line. The opening width and length are 80 nm and 100 nm, respectively. The scale bar is 500 nm. (b) A schematic cross-sectional view of this device.

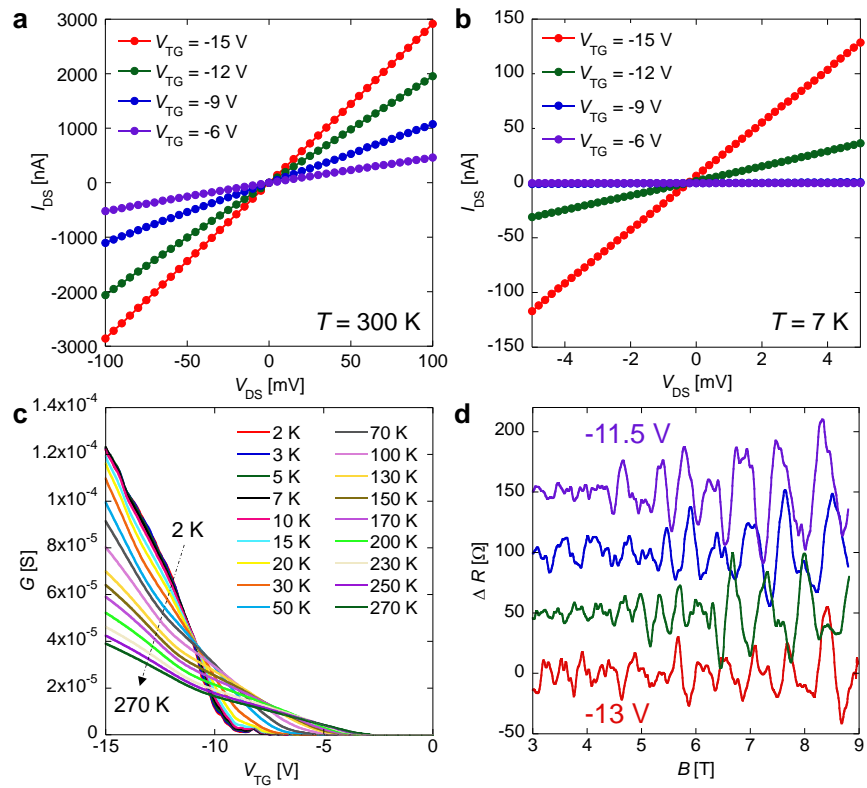

Figure 2. Two-terminal Current-Voltage plot at $V_{TG} = -6$ (purple), $-9$ (blue), $-12$ (green), and $-15$ V (red) at (a) $T = 300$ K and (b) $T = 7$ K. (c) Temperature dependent four terminal conductance curve as a function of $V_{TG}$. (d) Four-terminal resistance with subtraction of background resistance as a function of magnetic field at $T = 1.8$ K. $V_{TG} = -11.5, -12, -12.5,$ and $-13$ V from top to bottom.

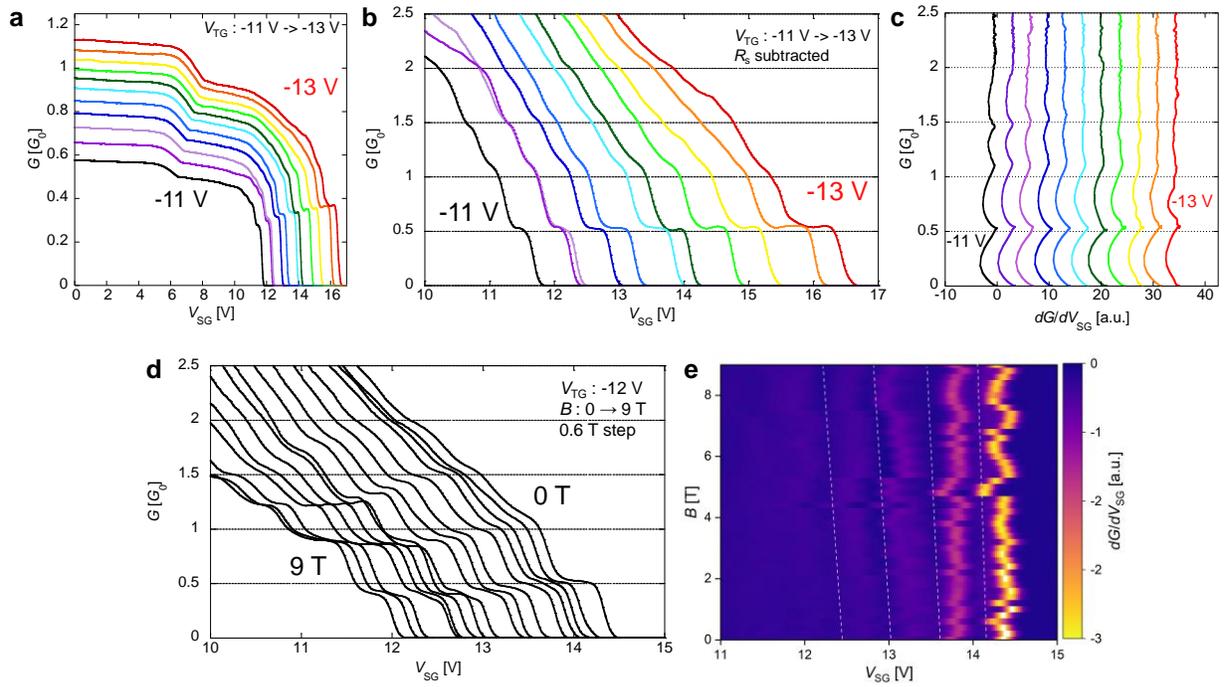

Figure 3. (a) Pristine conductance curves as a function of $V_{SG}$ at various $V_{TG}$ from −11 V to −13 V. The unit of $G_0$ is 2 $e^2/h$. (b) Conductance curves after correction by series resistance subtraction as $G = 1/(R_{4t} − R_S)$ where $R_s = R_{4t}(V_{SG} = 0)$. (c) Derivative of conductance curves of Figure 3b. The curves are offset along the horizontal direction for clarity. (d) Perpendicular magnetic field dependence of conductance curve as a function of $V_{SG}$ at constant $V_{TG} = −12$ V. Each curve is offset by −0.2 V along horizontal direction for visibility. (e) Color map of derivative of conductance $dG/dV_{SG}$ as function of magnetic field and split gate voltage $V_{SG}$.

(Full column)

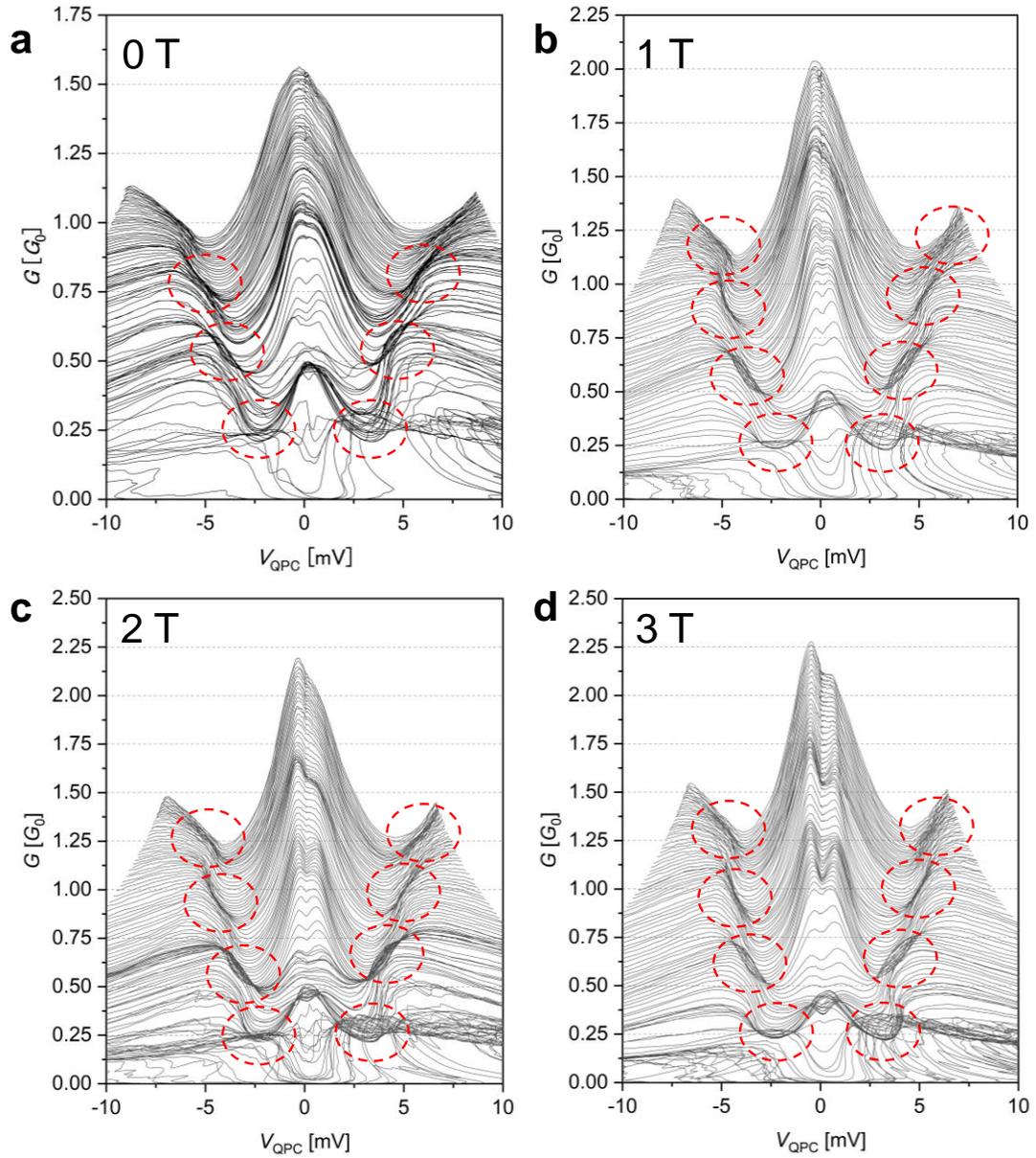

Figure 4. Nonlinear differential conductance ($G = dI/dV_{QPC}$) as a function of corrected bias voltage $V_{QPC}$ at constant $V_{TG}$ under $B =$ (a) 0, (b) 1, (c) 2, and (d) 3 T ranging from $V_{SG} =$ 11.5 V to 14 V. Red dashed circles in (a) are indicating quarter plateaus.

Supporting Information

# Signature of spin-resolved quantum point contact in p-type trilayer WSe$_2$ van der Waals heterostructure

Kohei Sakanashi,[1,*] Peter Krüger,[1] Kenji Watanabe,[2] Takashi Taniguchi,[3] Gil-Ho Kim,[4] David K. Ferry,[5] Jonathan P. Bird,[1,6] and Nobuyuki Aoki[1,*]

[1] Department of Materials Science, Chiba University, Chiba, 263-8522, Japan
[2] International Center for Materials Nanoarchitectonics, National Institute for Materials Science, Tsukuba, 305-0044, Japan
[3] Research Center for Functional Materials, National Institute for Materials Science, Tsukuba, 305-0044, Japan
[4] School of Electronic and Electrical Engineering and Sungkyunkwan Advanced Institute of Nanotechnology (SAINT), Sungkyunkwan University, Suwon, 16419, South Korea
[5] School of Electrical, Computer, and Energy Engineering, Arizona State University, Tempe, 85287, Arizona, USA
[6] Department of Electrical Engineering, the State University of New York, University at Buffalo, Buffalo, New York, 14260, USA

* To whom correspondence should be addressed: k.sakanashi@chiba-u.jp, n-aoki@faculty.chiba-u.jp

**Outline of supporting information**
**Section 1. Detailed device fabrication and measurement**
**Section 2. Thickness dependent of p$^+$-MoS$_2$ properties**
**Section 3. Global gate voltage dependence of conductance, resistance, and contact resistance**
**Section 4. Beating of SdH oscillations**
**Section 5. DFT calculations of trilayer WSe$_2$ under perpendicular electric field**
**Section 6. Detailed data analysis method of DC bias spectroscopy**
**Section 7. Temperature dependent QPC operation properties**
**Section 8. Differential conductance map of DC bias spectroscopy**

**S1. Detailed device fabrication and measurement**

All flakes of WSe$_2$ (flux grown crystals purchased from 2D semiconductor), hBN (NIMS), and p$^+$-MoS$_2$ (chemical vapor transport grown crystals purchased from HQ graphene) were exfoliated separately onto a solution and plasma cleaned 300 nm SiO$_2$/Si substrates. The thickness of those flakes was identified by optical contrast and atomic force microscopy (AFM). As we discuss later, the thickness of p$^+$-MoS$_2$ is quite important to achieve the low temperature Ohmic contact; we choose thick (> 30 nm) crystals.[1] The van der Waals heterostructure of

hBN/WSe$_2$/p$^+$-MoS$_2$/hBN were assembled by PC/PDMS based polymer based dry-transfer method.[2] To reduce the interface contamination, we use the dome-shaped PDMS (PDMS lens). The 8%w.t. PC thin film was placed on the top of the dome-shaped PDMS supported glass slide, and heated at 180 °C over 30 minutes to ensure the adhesion between PC film and PDMS lens/glass slide. The first hBN was picked up at 80 °C directly by PC film, and the other 2D crystal were picked up subsequently by using thermal- cycle van der Waals stamping method. The bottom split gates electrodes were prefabricated by means of electron beam lithography (EBL) with thin resist for higher resolution and electron beam deposition (EBD) of Cr 1 nm/Pd 9 nm. The cleanliness and flatness of bottom gates were identified by AFM and cleaned by means of sonification in Remover PG, acetone, IPA, and DI water. Then, substrate was annealed at 250 °C in high vacuum overnight. The source-drain electrodes were fabricated by EBL, reactive ion etching in a home-built etcher, and EBD of Cr 3 nm/Pd 15 nm/Au 100 nm to achieve Ohmic contact for p$^+$-MoS$_2$. The global gate electrode to tune the carrier density and contact properties were patterned by EBL and EBD on the top of the whole WSe$_2$ region. The device shows Ohmic contact behaviour at room temperature and low temperature, and the device properties have not degraded over 6 months.

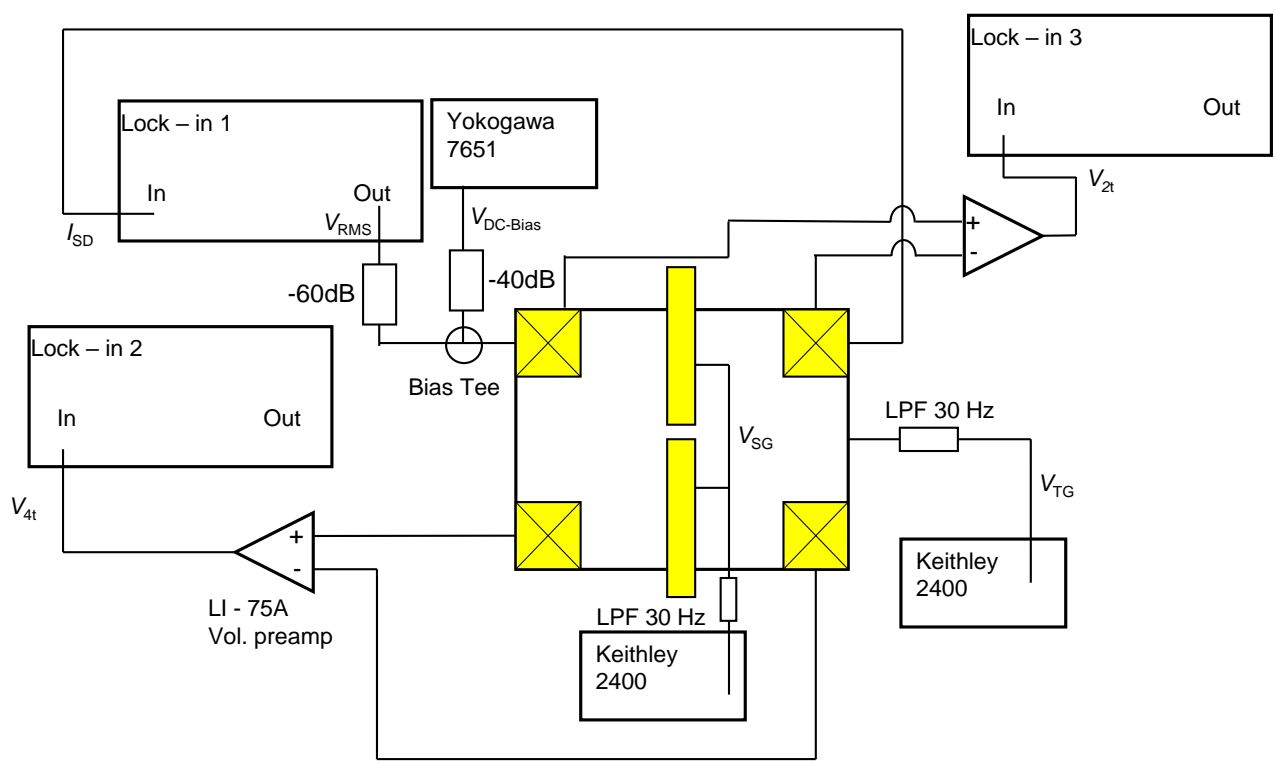

Figure S1. Schematic diagram of measurement set-up for WSe$_2$ quantum point contact.

All low temperature, measurements were carried out in a JANIS $^3$He insert at base temperature of 300 mK. The DC measurement lines were filtered by low pass filter, and current and voltage were measured by NF corporation LI5640 lock-in amplifiers. The gate voltage for the split gates and the top gate were supplied by Keithley 2400 source measure units, and DC bias voltage was applied by Yokogawa 7651. AC excitation from lock-in amplifier was attenuated to keep below 50 μV.

## S2. Thickness dependence of p$^+$-MoS$_2$ properties

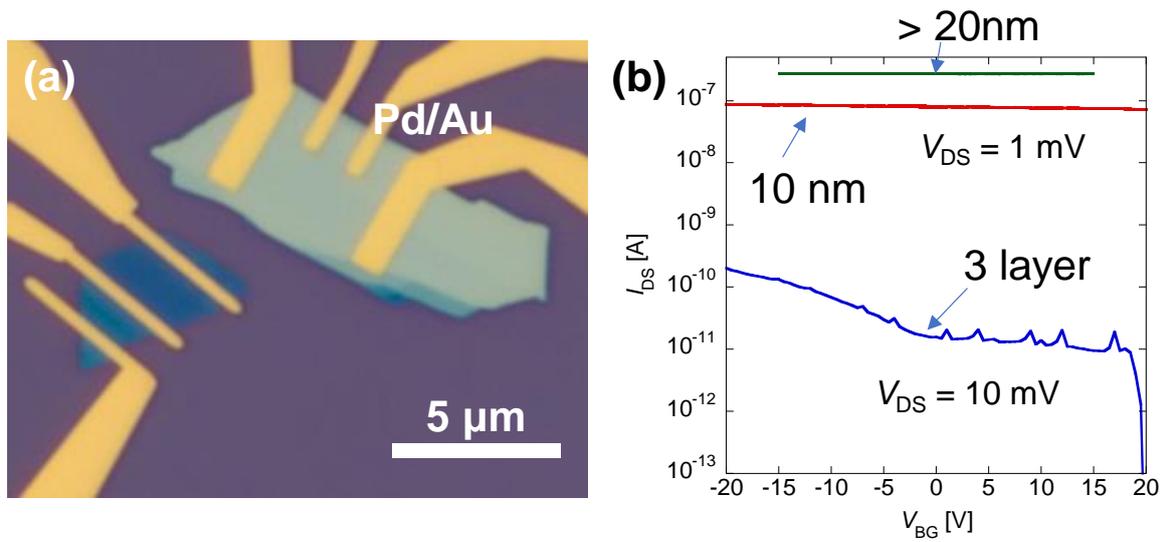

Figure S2. (a) The optical microscope image of p$^+$-MoS$_2$ devices. (b) The $I_{DS}$-$V_{BG}$ properties of various thickness of p$^+$-MoS$_2$ devices.

The key point to ensure the low temperature Ohmic contact for few layer WSe$_2$ is using thick (>30 nm) p$^+$-MoS$_2$ crystals as contact metals.[1,3] Figures S2(a, b) shows the typical optical image of p$^+$-MoS$_2$ devices and its gate voltage dependence of source-drain current, respectively. The p$^+$-MoS$_2$ purchased from HQ graphene company were grown using a chemical vapor transport method with small amounts of Nb atoms as dopant, and thin layers, or few layers, show n-type or ambipolar-like behavior, meaning not suitable for degenerately doped contacts. The thicker layers of 20 or 30 nm thick have no gate voltage dependence, and quite small resistance in kΩ range (See S3).

## S3. Global gate voltage dependence of conductance, resistance, contact resistance

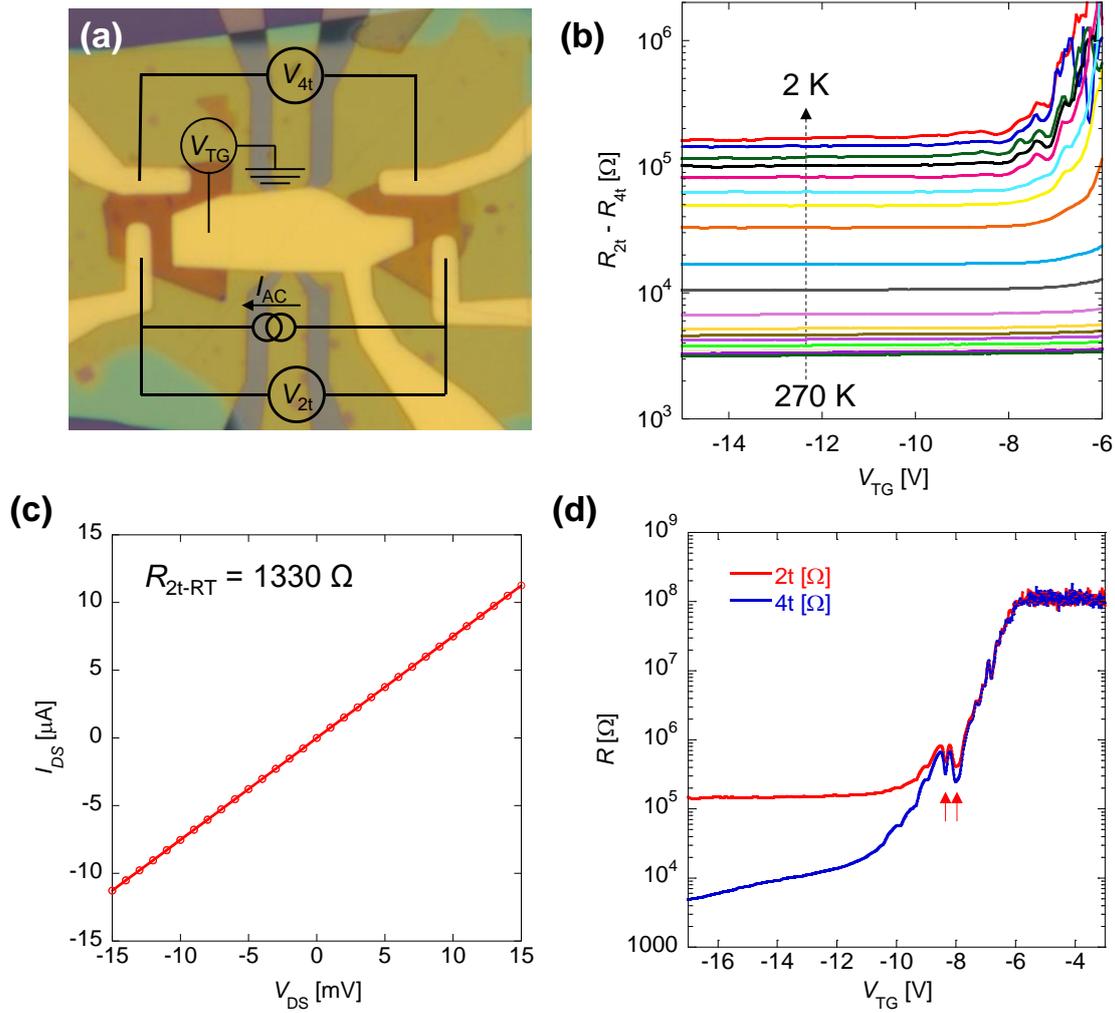

Figure S3. (a) Optical microscopy image of measured device and schematics measurement circuit model. The split gate electrodes for QPC measurement were grounded. (b) Temperature dependent subtraction of $R_{4t}$ from $R_{2t}$ as a function of $V_{TG}$. (c) The 2t I-V characteristics of $p^+$-$MoS_2$ at room temperature. (d) The $V_{TG}$ dependences of 2t resistance (red) and 4t resistance (blue). Both curves were measured simultaneously at 1.8 K. Red allow indicates the oscillatory behavior induced by localized states.

For this device structure, we need to determine carefully the absolute resistance value of the two-dimensional hole gas (2DHG) in order to correct the QPC properties by subtracting the series 2DHG resistance. Figure S3(a) schematically shows our quasi-4t measurement circuit. In this configuration, we can ignore the series line resistance, and contact resistance between the metal and $p^+$-$MoS_2$ as shown in Figure S3(b). However, the contact resistance of $p^+$-$MoS_2$/$WSe_2$ interfaces and the resistance of the two $p^+$-$MoS_2$ flakes around a few kilo-Ohms (Figure S3(c)) still remain. Then, the 4 terminal conductance discussed in this study shown in Figure 2a in main text are not equivalent to "true" 2DHG properties. The big differences between field effect mobility calculated by 4t conductance and quantum mobility calculated by starting point of Shubnikov-de Haas oscillations arises from this point. Even though this device shows better quantum mobility than previously reported devices, oscillatory behavior due to the localized state at lower carrier density was observed as shown in Figure S3(d). Due to the localized state caused by disorder of our sample, we tuned the carrier density above $3 \times 10^{12}$ cm$^{-2}$ and the region above the split gates are commonly completely depleted. And then the localization effect is suppressed. This localized state also affects the QPC channel region, however we fabricated multiple set of split gates, and chose the cleanest channel gates for measurement. This may be improved by using glovebox assembling or short LED flash at low temperature analogues to conventional HEMT devices.[4]

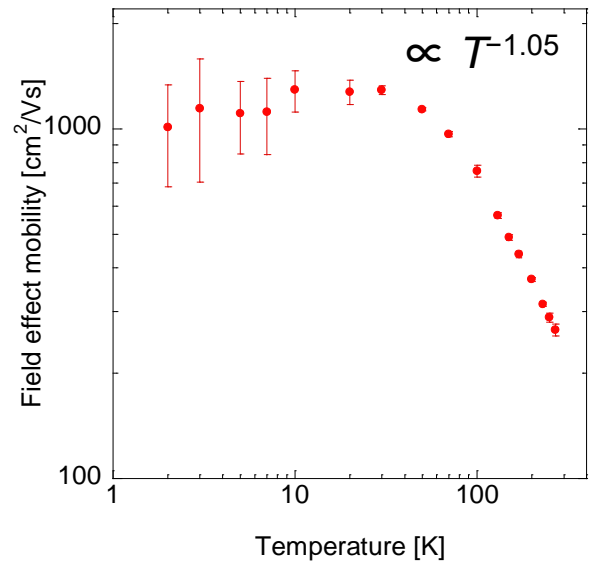

Figure S4. Temperature dependence of field effect mobility calculated by quasi-4t measurement.

## S4. Beating of SdH oscillations

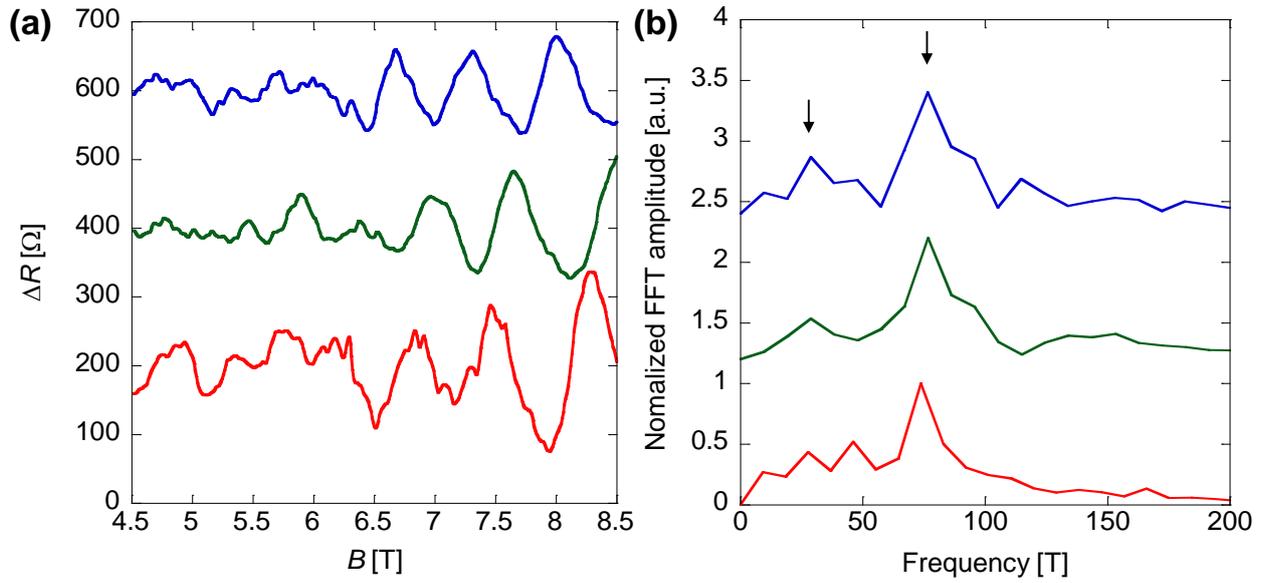

Figure S5. (a) Background subtracted magneto-resistance and (b) corresponding FFT curves for $V_{TG}$ = −11.5 V (red), −12.0 V (green), and −12.5 V (blue).

The SdH at particular carrier densities has beating of frequency as shown in Figure S5(a). From the $1/B$ FFT analysis of Figure S5(b), two components of oscillation were observed which is indicated by black allows. We think that this beating of SdH oscillations may be a result of a carrier density inhomogeneity due to charge impurity, strain effect, subband splitting of $\Gamma$ point, or doubling of K and $\Gamma$ point band edge.

## S5. DFT calculations of trilayer WSe$_2$ under perpendicular electric field

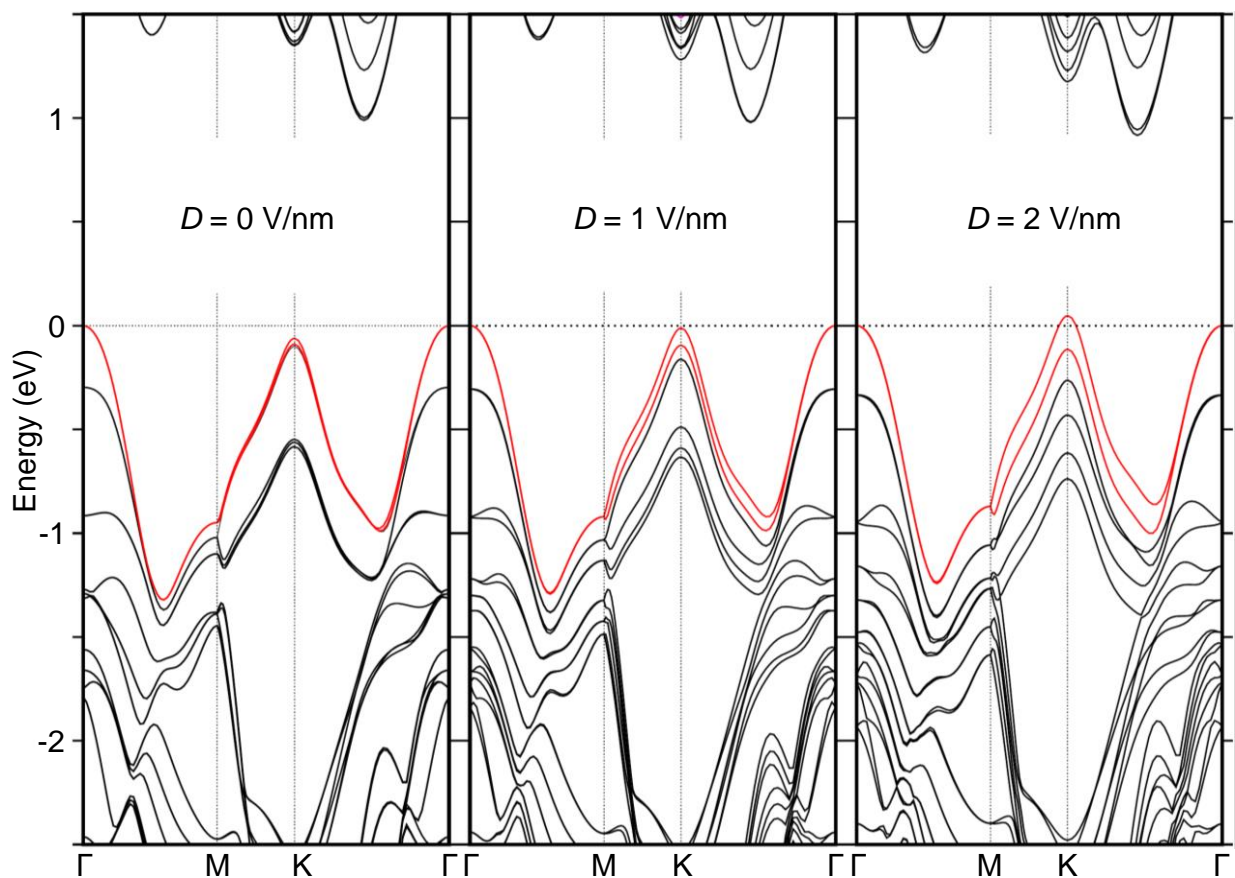

Figure S6. DFT calculated bands of trilayer WSe$_2$ under perpendicular electric field of 0, 1, and 2 V/nm (left to right). Band structure at the top of valence band are highlighted by red color for better visibility and traceability of lines.

Density functional theory (DFT) calculations were performed with the Vienna ab initio simulation package (VASP)[5] in the local density approximation with van-der-Waals (D2-Grimme) correction. The lattice constant of WSe$_2$ optimized in bulk are a = 3.330 Å and c = 12.789 Å which is within 1.5% of the experimental values. Slabs are separated by a vacuum of 15 Å, and spin-orbit coupling is included in all calculations. An external electric field is applied perpendicular to the trilayer by using the EFIELD feature in VASP together with dipole corrections[6] which cancel the spurious interaction between repeated slabs. Note that the trilayer is free-standing in vacuum so $\varepsilon_r = 1$ and $E = D$ which is the displacement field in main text. The calculated band structure, shown in Figure S6, reveals that the highest conduction band (in red) is spin-split by spin-orbit coupling and that the K-point becomes the top valence state for an electric field exceeding about 1 V/nm in agreement with previous reports.[7,8]

## S6. Detailed data analysis method of DC bias spectroscopy

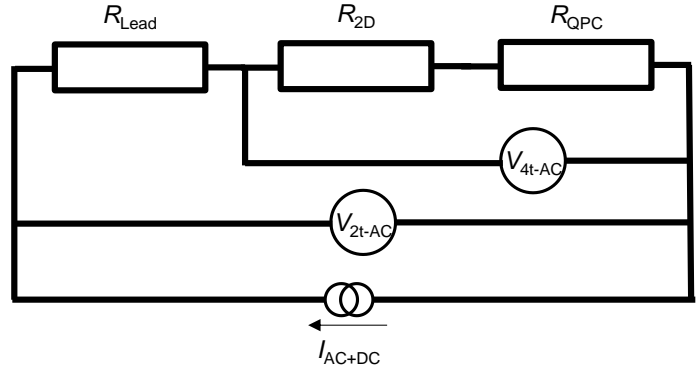

Figure S7. Schematic circuit model of QPC device with DC and AC bias voltage.

To estimate the exact value of true bias voltage at QPC region, we need to convert the applied bias voltage. We have a series contact resistance between the metal/p$^+$-MoS$_2$, a resistance of p$^+$-MoS$_2$, a contact resistance between p$^+$-MoS$_2$/WSe$_2$, and a resistance of WSe$_2$ 2DHG. We first define the series resistances as

$$R_{\text{Lead}} = R_{\text{filter+line}} + R_{\text{metal/p}^+-\text{MoS}_2}$$
$$R_{\text{2D}} = R_{\text{p}^+-\text{MoS}_2} + R_{\text{WSe}_2/\text{p}^+-\text{MoS}_2} + R_{\text{WSe}_2}.$$

Both of the series resistances can be measured by AC lock-in techniques with 2t and 4t voltage measurement simultaneously. The QPC resistance can be calculated as

$$R_{\text{QPC}} = R_{4t} - R_{4t(V_{\text{SG}}=0)}.$$

Then, the exact bias voltage at QPC can be calculated via simple calculation of voltage division alike

$$V_{\text{QPC}-\text{DC}} = V_{\text{SDbias}-\text{DC}} \times \frac{R_{4t} - R_{4t(V_{\text{SG}}=0)}}{R_{2t}}.$$

## S7. Temperature dependent QPC operation properties

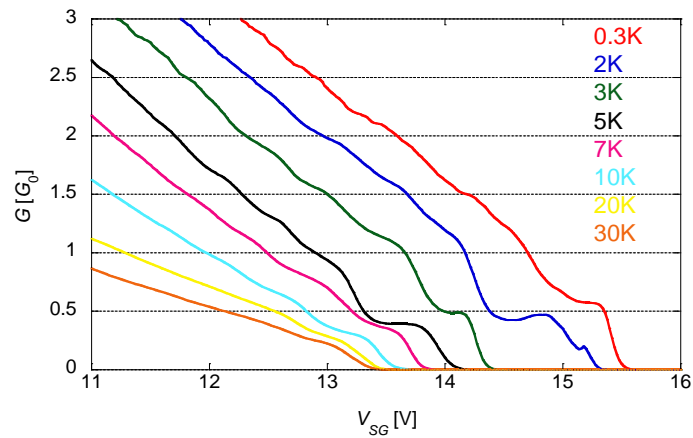

Figure S8. Temperature dependence of QPC properties at constant top gate voltage of $V_{\text{TG}} = -12$ V. All curves are corrected by subtracting series resistance. All curves are offset for clear visibility.

The temperature dependent QPC study reveals an estimation of the effective energy spacing of quantized energy levels in the electrostatically induced harmonic QPC potential. The QPC curves from 300 mK to 20 K show well-developed quantized plateaus at $e^2/h$ steps as shown in Figure S8. Lowering the background conductance by increasing the temperature causes the transition from ballistic transport to diffusive transport in QPC region by lowering hole mobility.

However, plateau-like kinks can be observed up to 30 K meaning that the energy spacing between each subband is approximately 2.6 meV which is comparable to the subband energy extracted by bias spectroscopy (See S8 for details). It is remarkable that this higher temperature QPC operation with relatively large subband energy spacing occurs, whereas the mobility is significantly smaller than in a conventional HEMT system.

## S8. Differential conductance map of DC bias spectroscopy

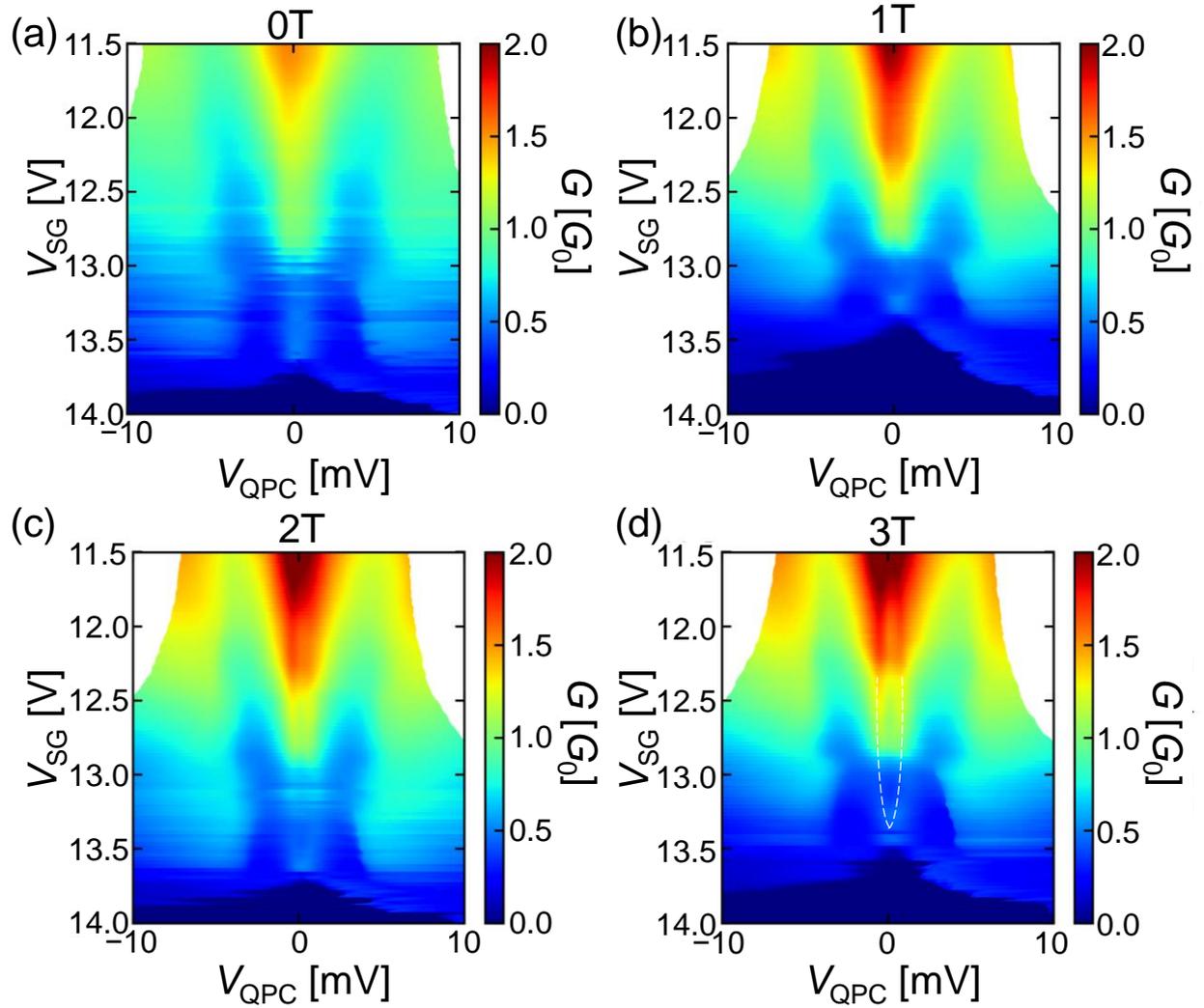

Figure S9. (a-d) Conductance color map as function of $V_{SG}$ and corrected bias voltage $V_{QPC}$ at perpendicular magnetic field of 0 - 4 T. White dashed line indicates the split peak positions.

Clear pinch-off states were observed for all magnetic field conditions. Segmented conductance have been seen for various bias voltages. At high magnetic field, the finite splitting of zero bias peaks have been observed as highlighted by white dotted line in Figure S9(d).

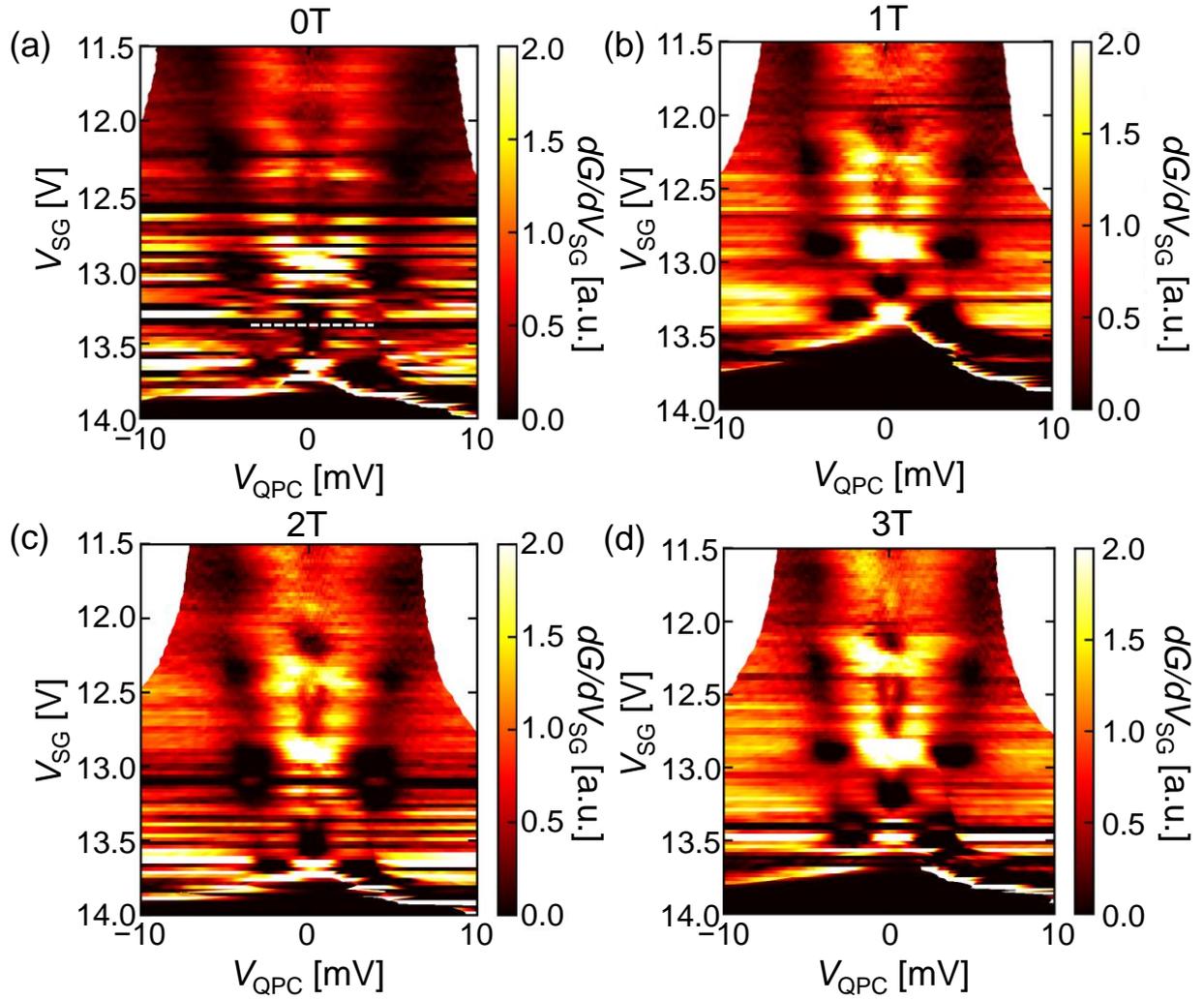

Figure S10. (a-d) Differential conductance color map as a function of $V_{SG}$ and corrected bias voltage $V_{QPC}$ at perpendicular magnetic field of 0 - 4 T. White dashed line in (a) indicates the bias voltage for next plateaus.

Transition line (bright spot) is unclear for every condition, however the plateaus (dark spot) is clear. The energy spacing between each subband can be calculated by $V_{SD} = 2\,\Delta_{SD}$ where $\Delta_{SD}$ is the subband energy spacing. The estimated subband energy for the first subband is approximately 2.7 meV.

# References

(1) Fang, N.; Toyoda, S.; Taniguchi, T.; Watanabe, K.; Nagashio, K. Full Energy Spectra of Interface State Densities for *n*- and *p*-type $MoS_2$ Field-Effect Transistors. *Adv. Funct. Mater.* **2019**, *29* (49), 1904465. https://doi.org/10.1002/adfm.201904465.

(2) Wang, L.; Meric, I.; Huang, P. Y.; Gao, Q.; Gao, Y.; Tran, H.; Taniguchi, T.; Watanabe, K.; Campos, L. M.; Muller, D. A.; Guo, J.; Kim, P.; Hone, J.; Shepard, K. L.; Dean, C. R. One-Dimensional Electrical Contact to a Two-Dimensional Material. *Science (80-. ).* **2013**, *342* (6158), 614–617. https://doi.org/10.1126/science.1244358.

(3) Takeyama, K.; Moriya, R.; Watanabe, K.; Masubuchi, S.; Taniguchi, T.; MacHida, T. Low-Temperature p-Type Ohmic Contact to $WSe_2$ using $P^+$-$MoS_2$/$WSe_2$ van Der Waals Interface. *Appl. Phys. Lett.* **2020**, *117* (15), 153101. https://doi.org/10.1063/5.0016468.

(4) Wang, K.; De Greve, K.; Jauregui, L. A.; Sushko, A.; High, A.; Zhou, Y.; Scuri, G.; Taniguchi, T.; Watanabe, K.; Lukin, M. D.; Park, H.; Kim, P. Electrical Control of Charged Carriers and Excitons in Atomically Thin Materials. *Nat. Nanotechnol.* **2018**, *13* (2), 128–132. https://doi.org/10.1038/s41565-017-0030-x.

(5) Kresse, G.; Furthmüller, J. Efficient Iterative Schemes for Ab Initio Total-Energy Calculations Using a Plane-Wave Basis Set. *Phys. Rev. B - Condens. Matter Mater. Phys.* **1996**, *54* (16), 11169–11186. https://doi.org/10.1103/PhysRevB.54.11169.

(6) Neugebauer, J.; Scheffler, M. Adsorbate-Substrate and Adsorbate-Adsorbate Interactions of Na and K Adlayers on Al(111). *Phys. Rev. B* **1992**, *46* (24), 16067–16080. https://doi.org/10.1103/PhysRevB.46.16067.

(7) Movva, H. C. P.; Lovorn, T.; Fallahazad, B.; Larentis, S.; Kim, K.; Taniguchi, T.; Watanabe, K.; Banerjee, S. K.; MacDonald, A. H.; Tutuc, E. Tunable Γ-K Valley Populations in Hole-Doped Trilayer $WSe_2$. *Phys. Rev. Lett.* **2018**, *120* (10), 107703. https://doi.org/10.1103/PhysRevLett.120.107703.

(8) Dai, X.; Li, W.; Wang, T.; Wang, X.; Zhai, C. Bandstructure Modulation of Two-Dimensional $WSe_2$ by Electric Field. *J. Appl. Phys.* **2015**, *117* (8), 084310. https://doi.org/10.1063/1.4907315.